\documentclass[aps,prd,preprint,groupedaddress]{revtex4-1}

\usepackage{amsmath}
\usepackage{graphicx}
\usepackage{axodraw4j}
\usepackage{slashed}

\begin{document}


\title{Cherenkov Radiation with Massive, CPT-violating Photons}


\author{Don Colladay and Patrick McDonald}
\affiliation{New College of Florida, Sarasota, FL, 34243}
\author{Robertus Potting}
\affiliation{CENTRA and Department of Physics, FCT, University of the Algarve,
8005-139 Faro, Portugal}


\date{\today}

\begin{abstract}
The source of CPT-violation in the photon sector of the Standard Model Extension
arises from a Chern-Simons-like contribution that involves a coupling
to a fixed background vector field $k_{AF}^\mu$.
These Lorentz- and CPT-violating photons have 
well-known theoretical issues that arise from missing states at low momenta 
when $k_{AF}^\mu$ is timelike.
In order to make the theory consistent,
a tiny mass for the photon can be introduced,
well below current experimental bounds.
The implementation of canonical quantization can then be implemented
as in the CPT-preserving case by using the St\"uckelberg mechanism.
We explicitly construct a covariant basis of properly-normalized
polarization vectors at fixed three-momentum
satisfying the momentum space field equations,
in terms of which the vector field can be expanded.
As an application of the theory,
we calculate the Cherenkov radiation rate for the case of
purely timelike $k_{AF}^\mu$,
and find a radiation rate at high energies that has a contribution that does not depend
on the mass used to regulate the photons.
\end{abstract}


\maketitle

\section{Introduction}

The Standard Model Extension (SME) is a framework that incorporates
Lorentz- and CPT-violating effects into the Standard Model \cite{kostelecky-colladay}.
Of particular interest is the pure gauge sector,
where Lorentz violation can be introduced,
either while preserving CPT or violating it.
Covariant quantization is extremely useful in performing
quantum-field-theoretic calculations as the formulas retain explicit
covariance throughout the computational procedure.
In a previous work we discussed how this can be implemented
for the CPT-preserving case \cite{colladay-mcdonald-potting1},
where it turned out that a consistent quantization requires the introduction
of a mass regulator.
In this work we extend the techniques developed in that work
to the CPT-violating case,
which is implemented using a Chern-Simons-like term parametrized by
a fixed background vector $k_{AF}^\mu$ \cite{carroll-field-jackiw}.
This term has received intensive attention in the literature as it can
arise as a radiative correction from the fermion sector in the
presence of an axial vector term \cite{jackiw-kostelecky}.
While quantization of Maxwell-Chern-Simons theory has been addressed
before in the literature \cite{adam-klinkhamer,andrianov},
attention was restricted to the (massless) case of purely spacelike
$k_{AF}^\mu$ in axial gauge.
A situation of physical interest is that of vacuum Cherenkov radiation.
This has been studied before, but for the case of spacelike $k_{AF}^\mu$
\cite{lehnert-potting,kaufhold-klinkhamer}.
However, there has been recent work on the case of purely timelike $k_{AF}^\mu$,
in the context of classical Maxwell-Chern-Simons theory \cite{altschul}.
As an application of the formalism developed in this work we will
calculate the Cherenkov radiation rate in field theory at tree level.

An argument could be made that these CPT-violating effects in the photon sector
are irrelevant, because they are bounded observationally to minute levels
\cite{carroll-field-jackiw}.
We take the point of view that it is important as a matter of principle
to establish whether these effects impede a rigorous quantization, and if not,
in what way the standard procedures have to be modified.
Also, while CPT-violating effects are strongly bounded in the photon sector,
this is not at all the case in the gluon sector.
(Of course we are dealing here with the non-abelian case,
but nevertheless the photon case,
if not directly applicable,
may offer important lessons there as well.)

As referred to above, a central element of our construction is
the introduction of a nonzero photon mass.
As it turns out,
the ultra-tight observational bound on $k_{AF}^\mu$ is actually many
orders of magnitude below the bound on the photon mass,
making it actually natural to allow for a nonzero value.
In the case of timelike $k_{AF}^\mu$,
it is not possible to take the massless limit in a consistent way.
For spacelike values, the massless limit is possible.
Nevertheless, even in this case the introduction of a photon mass
is a useful way to regulate infrared divergences that occur in
loop diagrams,
also in the context of Lorentz-violating effects \cite{cambiaso-lehnert-potting2}.
We note that the introduction of a photon mass in the context of the SME
has been studied in the presence of both CPT-preserving and
CPT-violating terms at the level of the equations of motion and
the propagator \cite{cambiaso-lehnert-potting}.

In section~\ref{eigenvectors}, we analyze the equations of motions
in momentum space,
and show that for the case of timelike $k_{AF}^\mu$ the introduction
of a nonzero mass parameter avoids a region of momentum space
with imaginary energies.
We set up a covariant basis of polarization vectors that solve
the momentum space equations of motions,
and satisfy a modified orthogonality condition.
In section~\ref{orthogonality}, an orthogonality relation is derived
for the eigenvectors,
that is used to define a consistent normalization.
The field operator is then quantized in terms of creation and annihilation
operators in section~\ref{quantization}.
In section~\ref{vacuum-Cerenkov}, we work out the rate of Cherenkov
radiation at tree level for the case of purely timelike $k_{AF}^\mu$.
Finally we present our conclusions.

\section{CPT-violating Photon sector of the SME}
\label{eigenvectors}

CPT violation in the photon sector of the minimal SME is given by the Lagrangian
\begin{equation}
\mathcal{L}_{A,k_{AF}} = -\frac14 F_{\mu\nu} F^{\mu\nu} +
\frac12 k_{AF}^\kappa \epsilon_{\kappa\lambda\mu\nu}A^\lambda F^{\mu\nu}\,.
\label{L-massless}
\end{equation}
where $k_{AF}^\mu$ is an arbitrary fixed background vector.
As is well known \cite{carroll-field-jackiw},
for timelike $k_{AF}^\mu$ the dispersion relation following from
(\ref{L-massless}) has a tachyonic character:
there is no covariant separation between positive and negative energy states
and there are (low) momenta for which there are no
corresponding real solutions for the energy,
signaling an unstable theory that does not permit a consistent quantization.

As was noted first in \cite{alfaro-cambiaso},
a way around this problem is to introduce a small mass term for the photon
through the St\"uckelberg mechanism.
The photon Lagrangian becomes
\begin{equation}
\mathcal{L}_A = -\frac14 F_{\mu\nu} F^{\mu\nu} +
\frac12 k_{AF}^\kappa \epsilon_{\kappa\lambda\mu\nu}A^\lambda F^{\mu\nu}  
+ \frac12 m_\gamma^2 A_\mu A^\mu - \frac1{2\xi}(\partial_\mu A^\mu)^2 
\label{lagrangian-A}
\end{equation}
where $\xi>0$ is a gauge parameter.

The equation of motion following from (\ref{lagrangian-A}) is
\begin{equation}
\left[(p^2-m_\gamma^2)\eta^{\mu\beta}-(1-\xi^{-1})p^\mu p^\beta
-2i\epsilon^{\mu\nu\alpha\beta}k_{AF\,\nu} p_\alpha\right]
e^{(\lambda)}_\beta(\vec p)=0
\label{eom}
\end{equation}
where $e^{(\lambda)}_\beta(\vec p)$ are the eigenvectors,
which we will take to be a function of the spacelike three momentum.
The full dispersion relation can be written
\begin{equation}
(p^2 - \xi m_\gamma^2)(p^2 - m_\gamma^2)
\left[ (p^2 - m_\gamma^2)^2 + 4 \left( k_{AF}^2 p^2 - (k_{AF} \cdot p)^2\right)\right] = 0\,.
\label{dispersion-relation}
\end{equation}
We see that there is an unphysical, gauge-dependent mode
satisfying the dispersion relation $p^2=\xi m_\gamma^2$ and a
physical mode with the conventional dispersion relation which we will denote
by $e^{(0)}$ and $e^{(3)}$, respectively.
They can be constructed from $p^\mu$ and $k_{AF}^\mu$ subspace.
For $m_\gamma > 0$, it is straightforward to check that
\begin{equation}
e^{(0)\,\mu}(\vec p) = N_0 p^\mu,\qquad
e^{(3)\,\mu}(\vec p)
= N_3\left(k_{AF}^\mu-\frac{p\cdot k_{AF}}{m_\gamma^2}p^\mu\right)
\qquad(m_\gamma > 0) \,.
\label{e03-massive}
\end{equation}
where $N_0$ and $N_3$ are normalization constants.
We see that $e^{(0)}$ is always timelike,
while $e^{(3)}$ is generally spacelike.

The two remaining, physical modes (denoted by $e^{(+)}$ and $e^{(-)}$)
which satisfy the perturbed dispersion relation are spacelike
and generally exhibit birefringent behavior.

In this work we will assume that $k_{AF}^\mu$ is timelike.
We can then always choose an observer frame
in which $\vec k_{AF} = 0$, $k_{AF}^0>0$.
The perturbed energies become
\begin{equation}
\bigl(p_0^{(\pm)}\bigr)^2
= (|\vec p\,| \pm k_{AF}^0)^2 + m_\gamma^2 - (k_{AF}^0)^2
= |\vec p\,|^2 + \tilde m_\pm^2,
\label{timelike}
\end{equation}
where we defined the momentum-dependent quantities
\begin{equation}
\tilde m_\pm^2 = m_\gamma^2 \pm 2k_{AF}^0|\vec p\,|\,.
\label{tilde-m_pm}
\end{equation}
We see from (\ref{timelike}) that problems arise
when $m_\gamma \sim k_{AF}^0$ are similar in magnitude.
In fact,
it can be easily seen that $\bigl(p_0^{(\pm)}\bigr)^2$ are always
non-negative provided $m_\gamma \geq k_{AF}^0$.
From now on, we will assume this to be the case.
An explicit expression for the modes $e^{(\pm)}$ is given by
\begin{equation}
e^{(\pm)}(\vec p)=
N_\pm\left(
\begin{array}{c}
0 \\
p^1 p^2 \mp i p^3 |\vec p\,|\\
-p_1^2 - p_3^2 \\
p^2 p^3 \pm i p^1 |\vec p\,|
\end{array}
\right)\qquad (k_{AF}\cdot k_{AF}>0,\quad \vec k_{AF}=0)
\label{e_pm-timelike}
\end{equation}
where $N_\pm$ are normalization constants.
Note that this expression is indeterminate when $\vec p = p_2 \hat e_2$, in 
which case an alternative expression can be easily found.

It is interesting to compare the current experimental bounds on these parameters.
The mass of the photon can be bounded through examination of large-scale
magnetic fields which would be have a perturbed structure if the photon had
a significant mass.
The particle data group quotes
\begin{equation}
m_\gamma < 1 \times 10^{-27}\,\mbox{GeV}.
\end{equation}
It may be possible to improve on this bound by nine orders of magnitude or so
by verifying certain properties of galactic magnetic fields,
but this result is still many orders of magnitude from the best bounds on $k_{AF}$
\begin{equation}
k_{AF}^0 \lesssim 10^{-43}\,\mbox{GeV},
\end{equation}
from cosmological searches for birefringence \cite{carroll-field-jackiw}.
This implies that if one wants to construct a phenomenologically viable model
for photons with CPT-violation,
it is possible to assume a nonzero mass that would be entirely consistent
with experimental observations.

It is easy to see from (\ref{timelike}) 
that one of the perturbed dispersion relations implies spacelike four-momenta,
while the other involves superluminal propagation,
for sufficiently large values of $|\vec p\,|$.
This fact is well known in the literature \cite{kostelecky-lehnert}.

\section{Orthogonality of the eigenvectors}
\label{orthogonality}

One implication of (\ref{eom}) is found by dotting with $p_\mu$,
which yields the condition 
\begin{equation}
(p^2-\xi m_\gamma^2)(e^{(\lambda)}\cdot p) = 0 \,,
\label{transversality}
\end{equation}
indicating that 
the physical states $\lambda=+,-,3$ for which $p^2\ne \xi m_\gamma^2$
must be transverse to momentum.

An orthogonality relation can be derived writing the equation of motion in the form
\begin{align}
&\left[ (p_0^{(\lambda)})^2  \eta^{\mu\nu}
-(1-\xi^{-1})\left((p_0^{(\lambda)})^2\delta_0^\mu\delta_0^\nu
+ p^i\delta_i^\mu p_0^{(\lambda)}\delta_0^\nu
+ p_0^{(\lambda)}\delta_0^\mu p^i\delta_i^\nu \right)
- 2i k_{AF}^{\kappa} \epsilon_{\kappa 0}{}^{\mu\nu} p_0^{(\lambda)} \right] 
e_\nu^{(\lambda)}(\vec p) \nonumber\\
&\qquad\qquad\qquad = \left[ \omega_p^2\eta^{\mu\nu} + (1-\xi^{-1})p^i\delta_i^\mu p^j\delta_j^\nu
- 2i k_{AF}^{\kappa}\epsilon_{\kappa j}{}^{\mu\nu} p^j \right]e_\nu^{(\lambda)}(\vec p),
\end{align}
with $\omega_p = \sqrt{\vec p^2 + m_\gamma^2}$.
Note that all of the dependency on $p_0$ has been moved to the left-hand side
of the equation.
This equation can then by multiplied by $e^{*(\lambda^\prime)}_\mu(\vec p)$
on the left and subtracted from the corresponding equation with
$\lambda \leftrightarrow \lambda^\prime$ switched,
which leaves the right-hand side of the equation equal to zero.
In the terms on the left-hand side a common factor
$ p_0^{(\lambda)}-p_0^{(\lambda^\prime)}$ can be extracted.
Consequently, if $ p_0^{(\lambda)} \ne p_0^{(\lambda^\prime)}$,
the remaining expression has to vanish.
The result is the following orthonormality relation 
\begin{align}
&e^{*(\lambda^\prime)}_\mu(\vec p)\Bigl[
\left( p_0^{(\lambda)}(\vec p) + p_0^{(\lambda^\prime)}(\vec p)\right)
\left(\eta^{\mu\nu}-(1-\xi^{-1})\delta_0^\mu \delta_0^\nu\right)\nonumber\\
&\qquad\qquad{}-(1-\xi^{-1})p^i(\delta_i^\mu\delta_0^\nu + \delta_0^\mu\delta_i^\nu)
-2 i k_{AF}^\kappa \epsilon_{\kappa 0}{}^{\mu\nu} \Bigr] e_\nu^{(\lambda)}(\vec p)
=  2 p_0^{(\lambda)} \eta_{\lambda \lambda^\prime},
\label{orthog1}
\end{align}
where the normalization of the polarization vectors are chosen to
match the conventional case, and we define $\eta^{00}=1$,
$\eta^{++}=\eta^{--}=\eta^{33}=-1$.
A similar relation holds for polarization vectors of opposite momenta
\begin{align}
&e^{(\lambda^\prime)}_\mu(-\vec p)\Bigl[
\left( p_0^{(\lambda)}(\vec p) - p_0^{(\lambda^\prime)}(-\vec p)\right)
\left(\eta^{\mu\nu}-(1-\xi^{-1})\delta_0^\mu \delta_0^\nu\right)\nonumber\\
&\qquad\qquad{}-(1-\xi^{-1})p^i(\delta_i^\mu\delta_0^\nu + \delta_0^\mu\delta_i^\nu)
-2 i k_{AF}^\kappa \epsilon_{\kappa 0}{}^{\mu\nu} \Bigr] e_\nu^{(\lambda)}(\vec p)
=  0\,.
\label{orthog2}
\end{align}
Note that there is no complex conjugate on the
left-side polarization vector in this relation.

The normalization factors introduced in (\ref{e03-massive}) that follow
from (\ref{orthog1}) are 
\begin{equation}
|N_0| = \frac{1}{m_\gamma},\qquad
|N_3| = \left(-k_{AF}\cdot k_{AF}
+ \frac{(p\cdot k_{AF})^2}{m_\gamma^2}\right)^{-1/2}\,.
\label{N03-massive}
\end{equation}

\section{Quantization}
\label{quantization}

The canonical momentum is computed in the usual way by taking derivatives of
the Lagrangian with respect to the time derivative of the $A$ fields 
\begin{equation}
\pi^j = F^{j0} + \epsilon^{0 j}{}_{k l} k_{AF}^k  A^l, \quad \pi^0
= - \xi^{-1}\partial_\mu A^\mu .
\end{equation}
Imposing equal-time canonical commutation rules
\begin{equation}
[A_\mu(t,\vec x), \pi^\nu(t,\vec y)] =
i\delta_\mu^\nu \delta^3(\vec x - \vec y),
\label{comrels0}
\end{equation}
along with
\begin{equation}
[A_\mu(t,\vec x), A_\nu(t,\vec y)] = [\pi^\mu(t,\vec x), \pi^\nu(t,\vec y)]=0 ,
\label{comrels1}
\end{equation}
implements the standard canonical quantization in a covariant manner as is
done in the conventional Gupta-Bleuler method.
From these definitions we find the commutation relations:
\begin{align}
[A^\mu(t,\vec x),\dot A^\nu(t,\vec y)]&=
[\dot A^\mu(t,\vec x),A^\nu(t,\vec y)]=
i\left(\eta^{\mu\nu}-\delta^\mu_0\delta^\nu_0(1-\xi)\right)
\delta^3(\vec x-\vec y)\,,
\label{comrels2}\\
[\dot A_\mu(t,\vec x),\dot A_\nu(t,\vec y)]&=
i\left[2\epsilon^{0\mu\nu\lambda}k_{AF\,\lambda}
+(1-\xi)\left(\delta^\mu_0\delta^\nu_j+\delta^\mu_j\delta^\nu_0\right)
\partial_x^j\right]\delta^3(\vec x-\vec y)\,. 
\label{comrels3}
\end{align}

Using the mode eigenvectors introduced in the previous section
the field can be expanded in terms of Fourier modes as
\begin{equation}
A_\mu(x) = \int \frac{d^3 \vec p}{(2 \pi)^3} \sum_\lambda
\frac{1}{2 p_0^{(\lambda)}}\left( a^\lambda(\vec p) 
 e_\mu^{(\lambda)}(\vec p)
e^{-i p \cdot x} +  a^{\lambda\dagger} (\vec p)  e_{\mu}^{*(\lambda)}(\vec p)
e^{i p \cdot x}\right).
\label{fouriermodes}
\end{equation}
This formula may be inverted using the orthogonality relations (\ref{orthog1}) and (\ref{orthog2}) as
\begin{align}
\eta_{\lambda\lambda'}a^{(\lambda')}(\vec q)&=
i\int d^3x e^{iq\cdot x}\Bigl[
{\stackrel{\leftrightarrow}{\partial_0}}
\left(\eta^{\mu\nu}-(1-\xi^{-1})\delta^\mu_0\delta^\nu_0\right)\nonumber\\
&\qquad\qquad{}-(1-\xi^{-1})q^j(\delta^\mu_j\delta^\nu_0+\delta^\mu_0\delta^\nu_j)
+2k_{AF\,\kappa}\epsilon^{\kappa0\mu\nu}
\Bigr]e_\nu^{*\,(\lambda)}(\vec q) A_\mu(x)\,,\\
\eta_{\lambda\lambda'}a^{\dagger\,(\lambda')}(\vec q)&=
-i\int d^3x e^{-iq\cdot x}\Bigl[
{\stackrel{\leftrightarrow}{\partial_0}}
\left(\eta^{\mu\nu}-(1-\xi^{-1})\delta^\mu_0\delta^\nu_0\right)\nonumber\\
&\qquad\qquad{}-(1-\xi^{-1})q^j(\delta^\mu_j\delta^\nu_0+\delta^\mu_0\delta^\nu_j)
+2k_{AF\,\kappa}\epsilon^{\kappa0\mu\nu}
\Bigr]e_\nu^{(\lambda)}(\vec q) A_\mu(x)\,.
\end{align}
With some algebra one shows that these relations, together with the
commutation relations (\ref{comrels1})-(\ref{comrels3}) imply that
the oscillators satisfy the usual commutation relations
\begin{align}
[a^{(\lambda)}(\vec p), a^{\dagger\,(\lambda^\prime)}(\vec q)] &=
- (2\pi)^3 2 p_0^{(\lambda)} \eta^{\lambda \lambda^\prime} \delta^3(\vec p - \vec q)\,,\\
[a^{(\lambda)}(\vec p), a^{(\lambda^\prime)}(\vec q)]
&=[a^{\dagger\,(\lambda)}(\vec p), a^{\dagger\,(\lambda^\prime)}(\vec q)] = 0\, .
\end{align}
Application of this algebra of mode operators to the coordinate-space
commutation relations  (\ref{comrels1}), (\ref{comrels2}), and (\ref{comrels3})
for the vector potential imply a variety of identities for bilinears 
of the polarization vectors including
\begin{align}
\sum_{\lambda,\lambda'}\eta_{\lambda,\lambda'}\frac{1}{p^{0\,(\lambda)}(\vec p)}
e_\mu^{(\lambda)}(\vec p)e_\mu^{*\,(\lambda')}(\vec p)
-
\sum_{\lambda,\lambda'}\eta_{\lambda,\lambda'}\frac{1}{p^{0\,(\lambda)}(-\vec p)}
e_\mu^{*\,(\lambda)}(-\vec p)e_\mu^{(\lambda')}(-\vec p)&=0\,,
\label{bilinear-identity-1}
\\
\frac12 \sum_{\lambda,\lambda^\prime} \eta_{\lambda \lambda^\prime}\left[ e_\mu^{(\lambda)}(\vec p) e_\nu^{*(\lambda^\prime)}(\vec p) 
+ e_\mu^{*(\lambda)}(-\vec p) e_\nu^{(\lambda^\prime)}(-\vec p)  \right] &=  \eta_{{\mu\nu}}\,.
\label{bilinear-identity-2}
\end{align}
Note that the right-hand side is independent of momentum indicating that
a complete set of polarization vectors must exist for every momentum choice. 
Unfortunately, this is not always true in the massless limit, 
so the presence of a mass term is crucial for the
consistency of the quantization procedure.

\section{Vacuum Cherenkov radiation}
\label{vacuum-Cerenkov}

It is well known that the deformed dispersion relations in the
Maxwell-Chern-Simons theory can give rise to vacuum Cherenkov radiation.
This has been worked out in detail in particular for the case of
spacelike $k_{AF}^\mu$,
both by analyzing the classical equations of motion
\cite{lehnert-potting} as well as in the context of quantum field theory
\cite{kaufhold-klinkhamer}.
The purely timelike case was considered very recently on the level of
the classical equations of motion \cite{altschul},
with the conclusion that the radiation rate is exactly zero.

We will use the formalism developed above to analyse this case
in the context of quantum field theory.
The relevant process is
\begin{equation}
e^- \to e^- + \gamma\,.
\end{equation}
We will consider this process at tree level,
represented by the Feynman diagram
\begin{center}
\begin{picture}(200,100)(0,0)
    \SetWidth{0.7}
    \SetColor{Black}
    \Photon[sep=4,clock](80,40)(120,80){2}{4.5}
    \Line[arrow,arrowpos=0.5,arrowlength=4,arrowwidth=3.5,arrowinset=0.2,sep=4](0,40)(80,40)
    \Line[arrow,arrowpos=0.5,arrowlength=4,arrowwidth=3.5,arrowinset=0.2,sep=4](80,40)(120,0)
    \Text(5,32)[cc]{\small $q$}
    \Text(128,76)[cc]{\small $p$}
    \Text(148,8)[cc]{\small $q'=q-p$}
\end{picture}
\end{center}
which we will analyze in the context of electrodynamics.
We assume for the lagrangian
\begin{equation}
\mathcal{L} = \mathcal{L}_A + \mathcal{L}_f
\label{lagrangian}
\end{equation}
with the standard Dirac fermion lagrangian
\begin{equation}
\mathcal{L}_f = \int d^4 x\,\bar\psi(i\slashed\partial-ie\slashed A + m)\psi
\end{equation}
which we assume to be without any Lorentz violation.

Cherenkov radiation is possible if there is a range of three-momenta for
which the fermion group velocity exceeds the photon phase velocity.
This in turn implies that the latter should be subluminal, that is,
the photon four-momentum should be spacelike.
This can only happen for the $e^{(-)}$ mode,
for sufficiently large three-momenta,
as we can see from (\ref{timelike}) and (\ref{tilde-m_pm}).

For the differential decay rate associated with with the Cherenkov radiation
process one finds
\begin{equation}
d\Gamma=
(2\pi)^4\frac{1}{2q^0}4m^2\frac{d^3\vec p}
{(2\pi)^3 p^0}\frac{d^3\vec{q'}}{(2\pi)^3(q')^0}
\delta^4(q'+p-q)\frac{1}{2}\sum_{\text{spins}}|\mathcal{A}|^2\,.
\label{decay-rate}
\end{equation}
Here we have averaged over the initial, and summed over the final
fermion spin states.
The fermion phase space factors in (\ref{decay-rate}) are conventional,
while the photon energy $p^0$ is,
of course, in accordance with the modified dispersion relation
(\ref{timelike}) corresponding to the $e^{(-)}$ mode.

More relevant than the differential decay rate
is the total radiation rate $W=-\dot{q}^0=\int p^0 d\Gamma$,
which becomes
\begin{align}
W &= \frac{m^2}{8\pi^2} \int \frac{d^3\vec{q'}\,d^3\vec p}{(q')^0\,q^0}
\,\delta^4(q'+p-q)\,\frac{1}{2}\sum_{\text{spins}}|\mathcal{A}|^2 \nonumber\\
&= \frac{m^2}{8\pi^2} \int \frac{d^3\vec p}{(q')^0\,q^0}
\delta\bigl((q')^0 + p^0 - q^0\bigr)\,
\frac{1}{2}\sum_{\text{spins}}|\mathcal{A}|^2\,.
\label{radiation-rate1}
\end{align}
For the matrix element we have
\begin{equation}
\mathcal{A} = ie\,\bar{u}(q') e^{(-)}_\mu(\vec p)\gamma^\mu u(q)\,.
\end{equation}
It follows that
\begin{align}
\frac{1}{2}\sum_{\text{spins}}|\mathcal{A}|^2
&= \frac{e^2}{2}\mbox{Tr}\left[\bar u(q)\gamma^\mu e^{(-)\,*}_\mu(\vec p) u(q-p)\,
\bar u(q-p)\gamma^\nu e^{(-)\,*}_\mu(\vec p) u(q) \right] \nonumber\\
&= \frac{e^2}{2m^2}\left(2q^\mu q^\nu - q^\mu p^\nu - p^\mu q^\nu
+ \tfrac{1}{2}\eta^{\mu\nu}\right)
e^{(-)\,*}_\mu(\vec p) \, e^{(-)}_\nu(\vec p)\,.
\label{spin-sum1}
\end{align}

Some algebra shows that
\begin{equation}
q^\mu q^\nu e_\mu^{(-)\,*}(\vec p)e_\nu^{(-)}(\vec p)
= \frac{1}{2} |\vec q|^2 \sin^2\theta  \,.
\label{qqee}
\end{equation}
where $\theta$ is the angle between $\vec q$ and $\vec p$, and Eq.(\ref{e_pm-timelike})
is used for $e_\mu^{(-)}$.
From the normalization condition (\ref{orthog1}) we find
\begin{equation}
\eta^{\mu\nu}e^{*\,(-)}_\mu(\vec p) e^{(-)}_\nu(\vec p) = -1\,.
\label{minus-norm}
\end{equation}
Using (\ref{spin-sum1}), (\ref{qqee}),  (\ref{minus-norm}),
as well as the orthogonality relation $p^\mu e_\mu^{(-)}(\vec p) = 0$ 
it then follows
\begin{align}
\frac{1}{2}\sum_{\text{spins}}|\mathcal{A}|^2
&= \frac{e^2}{2m^2}\left(|\vec q|^2 \sin^2\theta 
- \tfrac{1}{2}\tilde m_-^2\right)
\nonumber\\
&= \frac{e^2}{2m^2}\left(|\vec q|^2 \sin^2\theta - \tfrac{1}{2} m_\gamma^2
+ k_ {AF}^0|\vec p|\right)\,.
\label{spin-sum2}
\end{align}
Using this result in (\ref{radiation-rate1}) and introducing spherical coordinates
$
d^3\vec p = |\vec p|^2 d|\vec p|\,\sin\theta\,d\theta\,d\phi
$
one finds
\begin{align}
W &= \frac{e^2}{8\pi}\int\sin\theta\,d\theta\,d|\vec p|\,
\delta\bigl((q')^0 + p^0 - q^0\bigr)\frac{|\vec p|^2}{q^0(q')^0}
\left(|\vec q|^2 \sin^2\theta + k_ {AF}^0|\vec p| - \tfrac{1}{2}m_\gamma^2\right)
\nonumber\\
&= \frac{e^2}{8\pi q^0|\vec q|}\int_{p_{\text{min}}}^{p_{\text{max}}}
d|\vec p|\,|\vec p| \left(\bigl(k_ {AF}^0|\vec p| - \tfrac{1}{2}m_\gamma^2\bigr)
\left(\frac{2q^0(q^0-p^0) - k_ {AF}^0|\vec p| + \tfrac{1}{2}m_\gamma^2\bigr)}{|\vec p|^2} + 1 \right)
- m^2\right)
\label{radiation-rate2}
\end{align}
where
\begin{align}
p_{\text{max,min}}=\frac{1}{2\bigl(m^2 + 2k_ {AF}^0|\vec q|
- (k_ {AF}^0)^2\bigr)}&\biggl(2k_ {AF}^0(q^0)^2 - 2k_ {AF}^0 m_\gamma^2 + |\vec q|m_\gamma^2\nonumber\\
&\qquad{}\pm 2q^0\sqrt{(k_ {AF}^0q^0)^2 - k_ {AF}^0|\vec q|m_\gamma^2
- m^2m_\gamma^2+\tfrac{1}{4}m_\gamma^4}\biggr)  \,.
\label{p_max-min}
\end{align}
The limits $p_{\text{max,min}}$ in the integral over $|\vec p|$ arise
because the integral over $\theta$ in the second identity of (\ref{radiation-rate2})
requires that the delta-function representing energy conservation be
satisfied for some value of $\theta$.

Note also that the square root in the second term of (\ref{p_max-min})
imposes that its argument be non-negative.
This implies that
\begin{equation}
|\vec q| \ge \frac{m_\gamma^2 + 2m\sqrt{m_\gamma^2-(k_ {AF}^0)^2}}{2k_ {AF}^0} \equiv q_{\text{min}}
\label{condition_q_1}
\end{equation}
or
\begin{equation}
|\vec q| \le \frac{m_\gamma^2 - 2m\sqrt{m_\gamma^2-(k_ {AF}^0)^2}}{2k_ {AF}^0} \,.
\label{condition_q_2}
\end{equation}
As typically $m \gg m_\gamma > k_ {AF}^0$,
the right-hand side of (\ref{condition_q_2}) will be negative.
Therefore we will ignore the possibility that $\vec q$
satisfies condition (\ref{condition_q_2}).
We conclude that there is only a nonzero radiation rate
if the momentum of the incoming fermion exceeds the minimum value $q_{\text{min}}$.
For smaller momenta, the radiation rate is exactly zero.
A particularly interesting fact is the sensitive dependence of the threshold momentum
on the ratio $m_\gamma/k_ {AF}^0$.
For example, if this ratio is one, the threshold momentum is only of order $k_ {AF}^0$
which is surprising.

\begin{figure}
\includegraphics[scale=1]{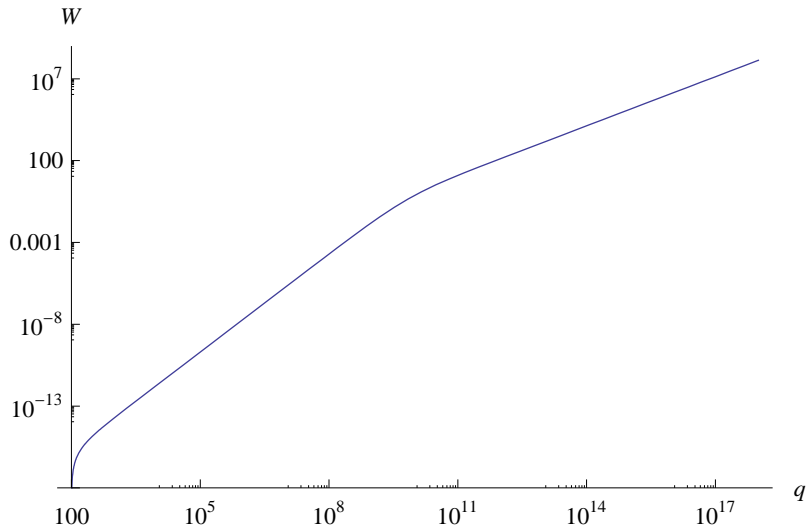}
\caption{Cherenkov radiation rate for purely timelike $k_{AF}$ as a
function of the fermion momentum.
Here we took the numerical values $m=1.0$, $k_ {AF}^0=1.0 \times 10^{-10}$,
$m_\gamma=1.0 \times 10^{-8}$, $e^2/(8\pi)=1$.}
\label{fig5}
\end{figure}

\begin{figure}
\includegraphics[scale=1]{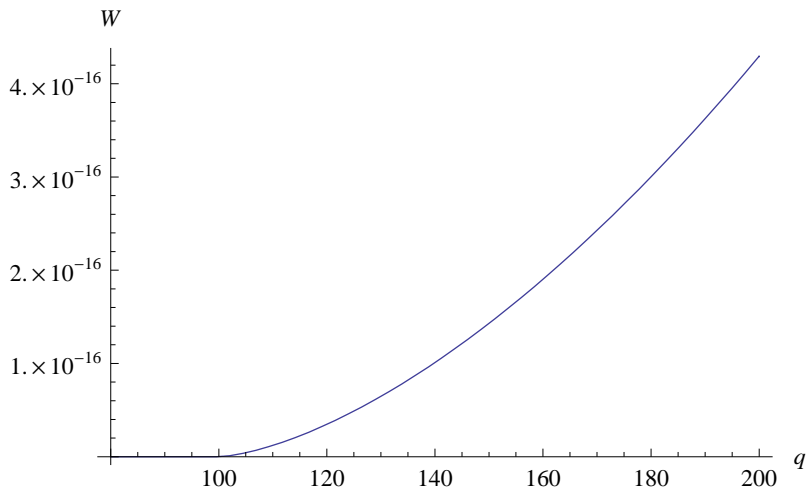}
\caption{Cherenkov radiation rate for purely timelike $k_{AF}$ for small
values of the fermion momentum.}
\label{fig6}
\end{figure}

The integral over $|\vec p|$ in (\ref{radiation-rate2}) can be evaluated
exactly with the use of Mathematica.
While the explicit result is not illuminating,
there are two asymptotic regimes for which the rate that can be
extracted from it:
\begin{align}
|\vec q| \gg \frac{m^2}{k_ {AF}^0}
&\qquad\Longrightarrow\qquad
W \sim \frac{e^2}{8\pi}\left(\frac{4k_ {AF}^0}{3}|\vec q| + \mathcal{O}(m^2)\right)
\label{regime1}\\
\frac{m\,m_\gamma}{k_ {AF}^0} \ll |\vec q| \ll \frac{m^2}{k_ {AF}^0}
&\qquad\Longrightarrow\qquad
W \sim \frac{e^2}{8\pi}\left(\frac{2(k_ {AF}^0)^2}{m^2}|\vec q|^2
+ \mathcal{O}(m_\gamma^2)\right) \,.
\label{regime2}
\end{align}
while
\begin{equation}
W=0\qquad \text{if}\qquad
|\vec q| < \frac{m\sqrt{m_\gamma^2-(k_ {AF}^0)^2}+\tfrac12 m_\gamma^2}{k_ {AF}^0}
\end{equation}
as we saw above.

In Figs.\ \ref{fig5} and \ref{fig6} the radiation rate is plotted for
particular values of the parameters.
The two regimes identified in (\ref{regime1}) and (\ref{regime2}) are clearly
visible in Fig.\ \ref{fig5}.
It is interesting that the (quadratic and linear) dependences on $|\vec q|$
in these regimes have analogues in the case of spacelike $k_{AF}^\mu$
\cite{lehnert-potting,kaufhold-klinkhamer}.
In addition, for the case $m_\gamma/k_ {AF}^0 \sim 1$, the very low-energy radiation rate is 
proportional to $(k_ {AF}^0)^2$ and is most likely unobservable.

Our work begins with the introduction of a nonzero mass
which insures well-defined real energies,
and, as we have shown, results in an observable physical effect:
the rate of Cherenkov radiation for purely timelike $k_{AF}^\mu$ is nonzero.
The literature includes attempts to treat the case of a massless photon
and the associated difficulties involving imaginary energies \cite{altschul}.
In particular,
the predictions involving the rate of Cherenkov radiation differ
from the nonzero mass case.

\section{Conclusions} 
\label{conclusions}

We have shown that a rigorous covariant quantization of
CPT-violating electrodynamics can be carried out for both timelike
and spacelike Lorentz-violating parameter $k_{AF}^\mu$.
In the former case it is necessary to introduce a photon mass
to avoid imaginary energies for small three momentum.
Our construction defines, at every value of three-momentum,
a basis of polarization vectors satisfying a modified orthogonality
relation.
The gauge potential can then be expanded consistently in terms of
creation and annihilation operators satisfying the usual commutation
relations.

We calculated the Cherenkov radiation rate in field theory at tree level,
obtaining an explicit expression for the rate.
The dependence of the rate on the fermion momentum is reminiscent of similar
dependences encountered in the case of spacelike $k_{AF}^\mu$.
A new, sensitive dependence of the threshold for radiation production
on the ratio of the photon mass to $k^0_{AF}$ is also found.

In this work we only considered the CPT-violating $k_{AF}$ parameter
of the minimal SME.
Recently, a generalization of this parameter has been introduced
involving higher-dimensional operators \cite{kostelecky-mewes}.
It would be interesting to investigate vacuum Cherenkov radiation
in this context.

\acknowledgments
We are thankful to Brett Altschul for discussions.
R.P.\ acknowledges financial support from the Portuguese Funda\c c\~ao
para a Ci\^encia e a Tecnologia.

\end{document}